\documentclass[prd,amssymb,10pt]{revtex4}
\usepackage{amsmath,amsfonts,amssymb}
\usepackage{verbatim,graphicx,float}

\begin{document}

\title{Relic gravitons as the observable for Loop Quantum Cosmology}

\author{Jakub Mielczarek}
\email{jakubm@poczta.onet.pl}
\affiliation{\it Astronomical Observatory, Jagiellonian University, 30-244
Krak\'ow, ul. Orla 171, Poland}

\author{Marek Szyd{\l}owski}
\email{uoszydlo@cyf-kr.edu.pl}
\affiliation{Astronomical Observatory, Jagiellonian University, 30-244
Krak\'ow, ul. Orla 171, Poland}
\affiliation{Marc Kac Complex Systems Research Centre, Jagiellonian University,
ul. Reymonta 4, 30-059 Krak{\'o}w, Poland}


\begin{abstract}
In this paper we investigate tensor modes of perturbations in the universe 
governed by Loop Quantum Cosmology.
We derive the equation for tensor modes
and investigate numerically effects of quantum corrections. 
This investigation reveals that the region of super-adiabatic 
amplification of tensor modes is smaller in comparison with
the classical case. Neglecting quantum corrections
to the equation for tensor modes and holding underlying loop 
dynamics we study analytically the creation of gravitons.  
We calculate the power spectrum of tensor perturbations during the 
super-inflationary phase induced by Loop Quantum Gravity. The main 
result obtained is the spectrum of gravitons, produced 
in the transition from the quantum to classical regime of the Universe. 
Obtained spectrum is characterized by a hard branch.
The numerical investigation shows the strong dependence of $\nu_{\text{max}}$
frequency with respect to quantum numbers. 
The parameter $\Omega_{\text{gw}}(\nu_{\text{max}})$ 
approaches even to $\sim 10^{-1}$ for highest frequencies. We compare our results with recent constraints 
from the Laser Interferometer Gravitational-wave Observatory (LIGO) and find 
that it is possible to test the quantum effects in the early Universe.
\end{abstract}

\maketitle

\section{Introduction}
Loop Quantum Gravity (LQG) introduces strong modifications to the standard 
description of the early universe. The main difference with the classical 
approach is the avoidance of an initial singularity \cite{Bojowald:2001xe}. 
This effect leads to the bouncing solution on the semi-classical level 
\cite{Bojowald:2005zk,Stachowiak:2006uh}. Another interesting property is the 
occurrence of the super-inflationary phase induced by quantum effects 
\cite{Bojowald:2002nz}. This phase is in fact not long enough to explain the 
observed flatness of the Universe, but after this phase the Universe has proper 
initial conditions to start the standard slow-roll inflation. In this scenario 
a inflaton field firstly climbs up the potential hill and then stops before a 
slow-roll phase, producing the running of the spectral index 
\cite{Tsujikawa:2003vr}. The production of scalar perturbations during 
super-inflationary phase is investigated in the papers 
\cite{Hossain:2005,Calcagni:2007,Mulryne:2006}.

In this paper we consider the transition from the quantum to classical universe 
through the super-inflationary phase. Because during this transition the main 
contribution to the energy of the inflaton field comes from the kinetic part, 
in calculations, we neglect the contribution from the potential energy. It is 
worthwhile to note here that the super-inflationary phase inducted by LQG is a 
generic property and does not depend on a kind of the field which fills the 
universe. For analytical considerations we solve dynamical equations in the
semi-classical and classical regimes and then we match them. It is done for
the value of the scale factor
\begin{equation}
a_0=a_*=\sqrt{\frac{\gamma j}{3}} l_{\text{Pl}}
\end{equation}                 
where $j$ is a half-integer quantization parameter, $l_{\text{Pl}}$ is a Planck 
length and $\gamma$ is the Barbero-Immirzi parameter. The latter parameter
\begin{equation}
\gamma =\frac{\ln 2}{\pi \sqrt{3}}
\end{equation}
comes from calculations of black-holes entropy \cite{Ashtekar:1997yu}. Below 
the value $a_*$ non-perturbative modifications become important. We derive
the equation for tensor modes in the LQG scenario. We investigate  numerically 
effects of loop corrections. Neglecting quantum corrections
to equation for tensor modes and holding underlying loop 
dynamics we study analytically creation of gravitons. We calculate 
the spectrum of tensor perturbations during the super-inflationary phase and 
the density of gravitons produced during the transition from the semi-classical 
to classical universe. Such gravitons give contributions to the stochastic 
background of gravitational waves. Nowadays the detectors like LIGO 
\cite{Abbott:2003vs} aim at the detection of these stochastic gravitational 
waves \cite{Abbott:2007wd}. Usually to describe the spectrum of gravitational 
waves the parameter    
\begin{equation}
\Omega_{\text{gw}}(\nu) =\frac{\nu}{\rho_c}\frac{d \rho_{\text{gw}}}{d \nu}
\label{omegaGW}
\end{equation}
is introduced. Here $\rho_c$ is the current critical density, $\rho_{\text{gw}}$ 
is the density of gravitational waves and $\nu$ is the physical frequency 
measured today. The recent LIGO constraint for this parameter is 
$\Omega_{\text{gw}} < 6.5 \cdot 10^{-5}$ \cite{Abbott:2006zx}. 
We also calculate the value of the function $\Omega_{\text{gw}}(\nu)$ in the 
model and compare it with LIGO constraints.

\section{The semi-classical dynamics}

\begin{figure}[ht!]
\centering
\includegraphics[width=7cm,angle=270]{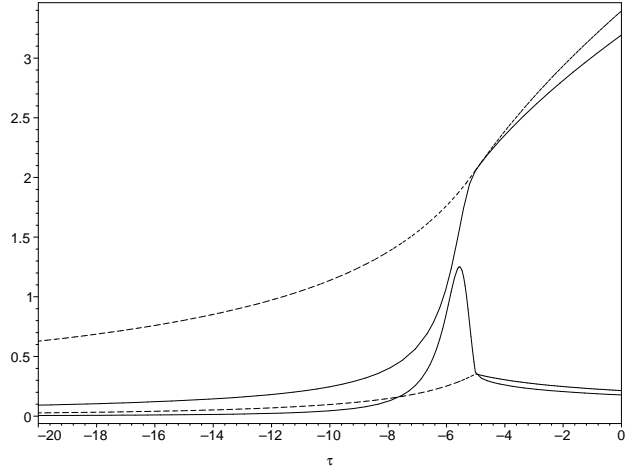}
\caption{The evolution of the scale factor $a$ (upper curve) and  $a'$ (lower curve) in
the conformal time  (with $\tau_0=5$, $j=100$ and $l=0.1$). The dashed line presents the 
approximate solution and the solid line corresponds to the numerical solution. The initial 
conditions are fixed for $a_*$.}
\label{dvolution}
\end{figure}

Loop Quantum Gravity introduce strong modifications to the dynamical equations 
in the semi-classical regime. These modifications come from the expression for 
the density operator \cite{Bojowald:2006da} 
\begin{equation}
d_j(a)=D(q)\frac{1}{a^3} 
\end{equation} 
where $q$ is defined as follow
\begin{equation}
q \equiv \frac{a^2}{a^2_*}
\end{equation}
and for the semi-classical universe ($l_{\text{Pl}} < a \ll a_*$) the quantum 
correction factor has a form \cite{Bojowald:2004ax}
\begin{equation}
D(q)=q^{3/2} \left\{ \frac{3}{2l} \left(  \frac{1}{l+2}\left[(q+1)^{l+2}-|q-1|^{l+2} \right]-
\frac{q}{1+l}\left[(q+1)^{l+1}-\mbox{sgn}(q-1)|q-1|^{l+1} \right]  \right) \right\}^{3/(2-2l)}.
\label{correction}
\end{equation}
Here $l$ is the ambiguous parameter of quantization constrained by $0<l<1$ \cite{Bojowald:2002ny}.
The Hamiltonian for the scalar field in the flat FRW universe has a form
\begin{equation}
\mathcal{H} = \frac{1}{2} d_j(a) p^{2}_{\phi}+a^3V(\phi) \ \ \mbox{where} \ \ p_{\phi}=d^{-1}_{j}(a)\dot{\phi} \ .
\end{equation} 
This lead to the equation of motion of the form 
\begin{equation}
\ddot{\phi}+\left(3H -\frac{\dot{D}}{D} \right)\dot{\phi} +D\frac{dV}{d\phi} =  0   \ .
\label{eom}
\end{equation}
The Friedmann and Raychaudhuri equations for the universe filled with a scalar 
field are respectively
\begin{eqnarray}
H^2 &=& \frac{8\pi G}{3} \left[ \frac{\dot{\phi}^2}{2D} +V(\phi)    \right]  \ ,    \label{Fried1}      \\
\frac{\ddot{a}}{a} &=& -\frac{8\pi G}{3} \left[ \frac{\dot{\phi}^2}{D} \left( 1-\frac{\dot{D}}{4HD}  \right) -V(\phi) \right].   
\end{eqnarray}
From equations (\ref{eom}) and (\ref{Fried1}) we obtain the relation   
\begin{equation}
\dot{H} = -4\pi G \frac{\dot{\phi}^2}{D} \left(1-\frac{\dot{D}}{6 H D}  \right).
\label{Hdot}
\end{equation}
Due to quantum correction $D$ in the region ($l_{\text{Pl}} < a \ll a_*$), the 
expression in the bracket can be negative, leading to 
$\dot{H}>0$ (super-inflation). If $a \gg  a_*$ then $D \approx 1 $ 
leading to  $\dot{H}<0$ (deceleration).   
For $a \ll  a_* $ the approximation of expression (\ref{correction})  have a form  
\begin{equation}
D(q) \approx \left( \frac{3}{1+l}  \right)^{3/(2-2l)} \left( \frac{a}{a_*}\right)^{3(2-l)/(1-l)} \ .
\end{equation}
We use this approximation to calculate the dynamics in the semi-classical region. Now  
\begin{equation}
\frac{\dot{D}}{H D} = \frac{3(2-l)}{1-l} \ > 6
\label{DHD}
\end{equation}
leading to the phase of acceleration, see equation (\ref{Hdot}).  
Putting (\ref{DHD}) into equation (\ref{Hdot}) and combining with (\ref{Fried1}) 
we obtain the equation for the scale factor  
\begin{equation}
a a'' -(a')^2 \left[2+\frac{3}{2}\frac{l}{1-l} \right]=0 
\label{equat1}
\end{equation}
where prime means the derivative in respect to the conformal time $d\tau=dt/a$. 
We assume here $V(\phi)=0$ as it was mentioned in section {\bf I}. The 
solution of (\ref{equat1}) is of the form 
\begin{equation}
a \propto (-\tau)^{-2\frac{1-l}{2+l}}.
\end{equation}
To calculate the solution in the classical regime we take $D=1$. In this case 
the equation for the scale factor have a form
\begin{equation}
a a'' +(a')^2=0. 
\end{equation}
Now we  match two solutions, from two regions, at some $\tau_0$ as follow
\begin{eqnarray}
a_1(-\tau_0) &=&  a_2(-\tau_0) \ , \\
a_1'(-\tau_0) &=&  a_2'(-\tau_0).
\end{eqnarray}
Where region 2 is classical and region 1 is semi-classical. The value of the 
chosen conformal time $\tau_0$ corresponds to the scale
factor $a_*$. After matching we obtain the solution of the form 
\begin{eqnarray}
a_1(\tau) &=&  a_* \left(-\frac{\tau}{\tau_0}\right)^{-2\frac{1-l}{2+l}}  \ \ \mbox{for}   \ \  \tau<-\tau_0  \ , \label{evol1}  
  \\
a_2(\tau) &=&  a_* \sqrt{4\frac{1-l}{2+l}\left(\frac{\tau_0+\tau}{\tau_0} \right) +1  }  \ \ \mbox{for}    \ \  \tau>-\tau_0  
 \ \     . \label{evol2}
\end{eqnarray}
This solution is shown in Fig.~\ref{dvolution} together with the numerical solution. 
The upper curve corresponds to the evolution of the scale factor $a(\tau)$, while 
the bottom curve does to the first derivative of the scale factor in respect to the conformal time.    
The obtained solution is of course only an approximation of the real evolution, 
however it is sufficiently exact to be used in analytical calculations. We find the 
agreement with the numerically calculated evolution of the scale factor and 
the Hubble rate obtained by Tsujikawa et al. \cite{Tsujikawa:2003vr}. 
In the future investigations we use both numerical and approximate analytical 
solutions. 

\section{Evolution of tensor modes in Loop Quantum Cosmology}

Tensor perturbations $h_{ij}$ to the FRW metric we can express as   
\begin{equation}
ds^2=a^2(\tau) \left[ -d\tau^2 +(\delta_{ij}+h_{ij} )dx^i dx^j \right]
\end{equation}
with $|h_{ij}|\ll 1$. Using constraints $h^i_i=\nabla_ih^i_j=0 $ we can see 
that tensor $h_{ij}$ have only two independent components $h^1_1=-h^2_2=h_+$ 
and $h^2_1=h^1_2=h_{\times}$. These components correspond to two polarizations 
of gravitational waves. Since tensor modes of perturbation are not coupled to 
the scalar field source, we can obtain equations for them from the variation 
of the action
\begin{equation}
S^{(2)}_t=\frac{1}{64\pi G} \int d^4x a^3 \left[ 
\partial_t h^i_j\partial_t h^j_i-\frac{1}{a^2}\nabla_k h^i_j\nabla_k h^j_i   \right]
 = \frac{1}{32\pi G} \int d^4x a^3 \left[ \dot{h}_{\times}^2+\dot{h}_{+}^2-
\frac{1}{a^2}\left(\vec{\nabla} h_{\times} \right)^2-\frac{1}{a^2}\left(\vec{\nabla} h_{+}\right)^2  \right].
\label{action1}
\end{equation}
For the detailed discussion of this issue see Ref.~\cite{Giovannini:2004rj,Giovannini:2007xh}.
Both polarizations of gravitational waves are not coupled and can be treated separately. Introducing
\begin{equation}
h=\frac{h_{+}}{\sqrt{16\pi G}}=\frac{h_{\times}}{\sqrt{16\pi G}}
\end{equation}
we can rewrite the action for tensor modes in the form
\begin{equation}
S^{(2)}_t = \frac{1}{2} \int d^4 a^3 \left[\dot{h}^2-\frac{1}{a^2} \left(\vec{\nabla} h \right)^2  \right] .
\label{action2}
\end{equation}
Quantum corrections can be introduced now in the same way as 
in the scalar field case \cite{Bojowald:2006da,Bojowald:2004ax}, leading
to the equation of motion 
\begin{equation}
\ddot{h}+\left(3H-\frac{\dot{D}}{D} \right)\dot{h} -D\frac{\nabla^2 h}{a^2} = 0.
\label{tensormodes1}
\end{equation}
There are no other corrections from Loop Quantum Gravity which contribute to 
this equation.

\begin{figure}[ht!]
\centering
$\begin{array}{cc}   
\includegraphics[width=6cm,angle=270]{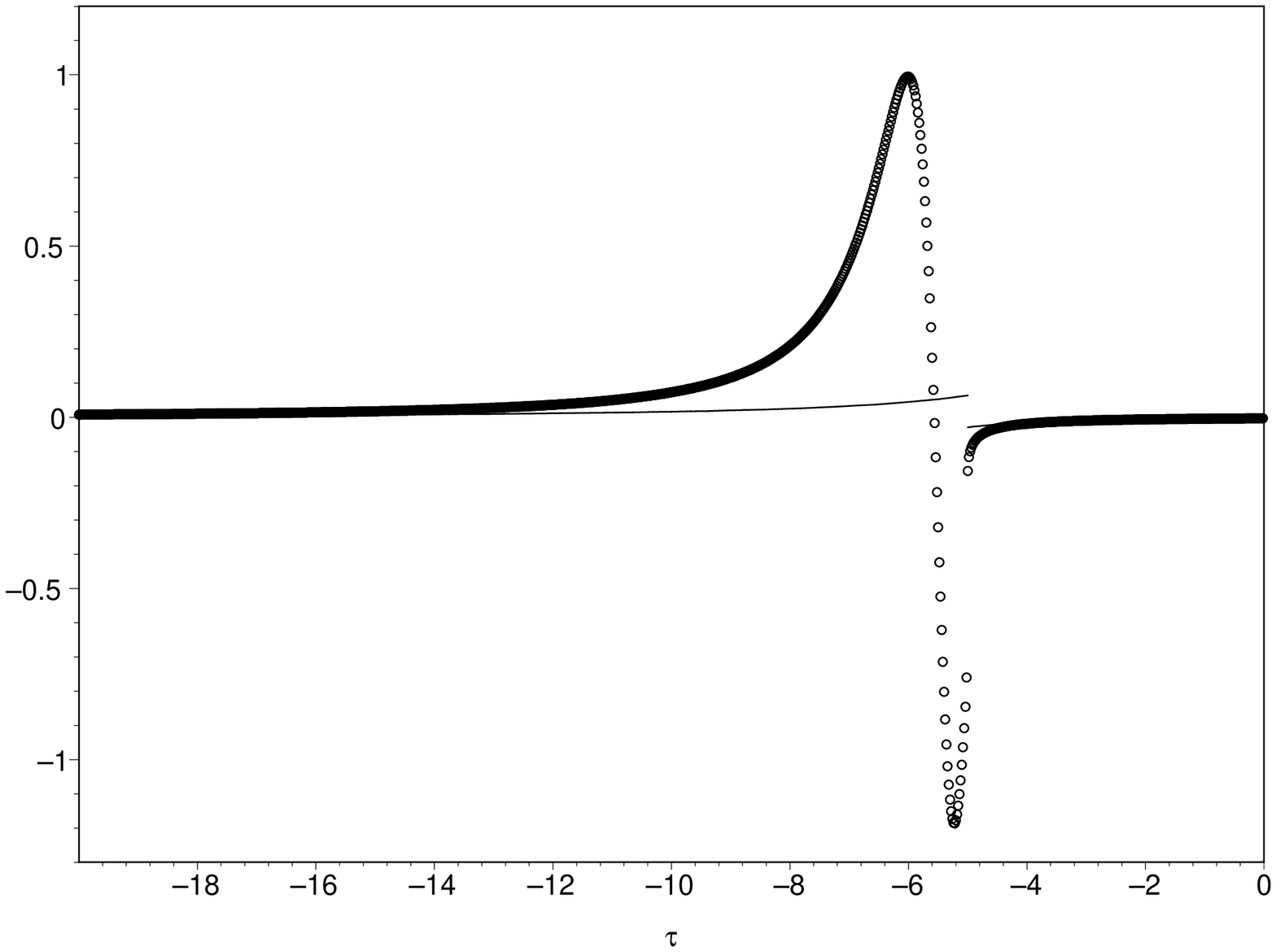}  &  \includegraphics[width=6cm,angle=270]{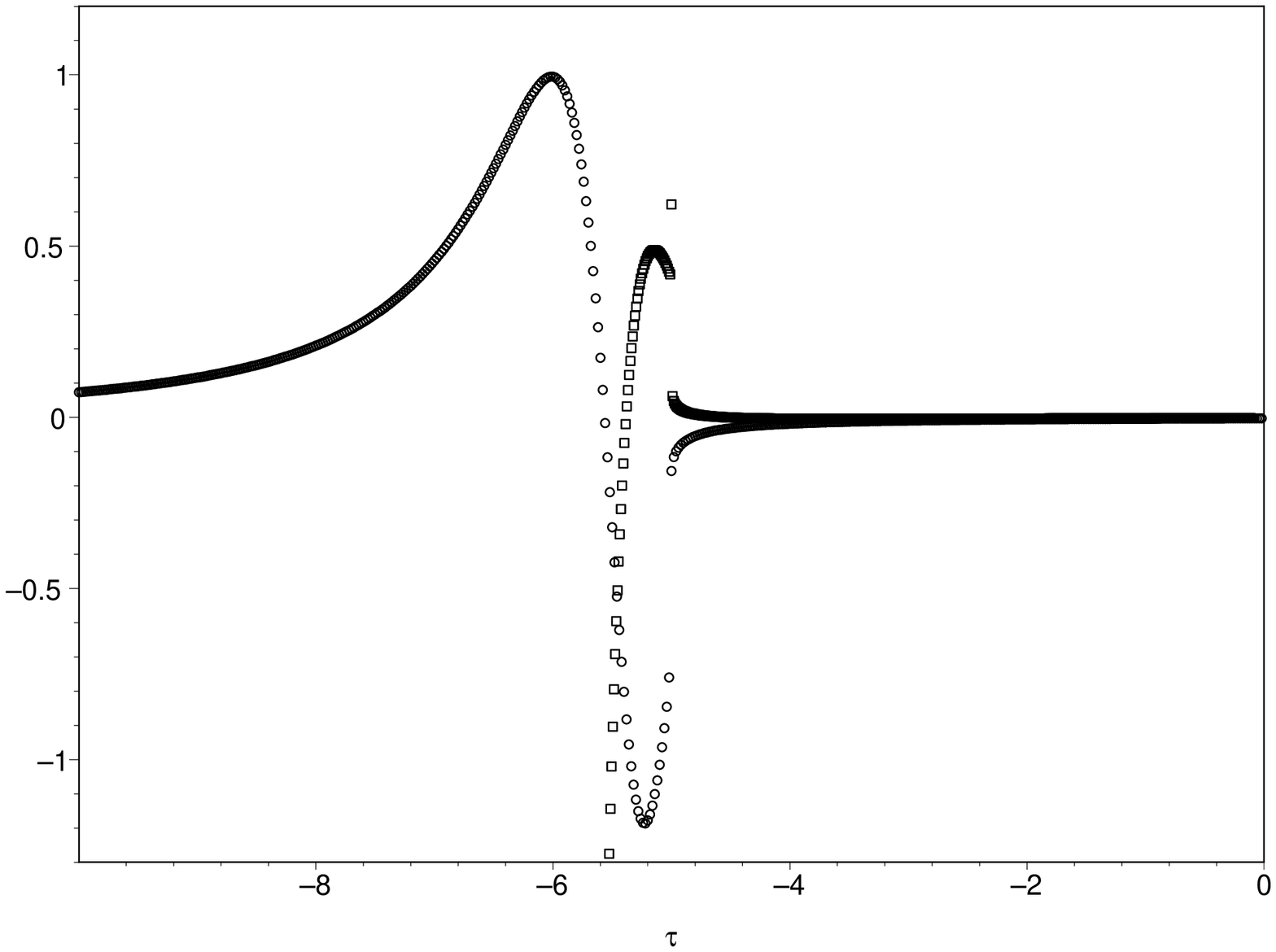}   \\
\includegraphics[width=6cm,angle=270]{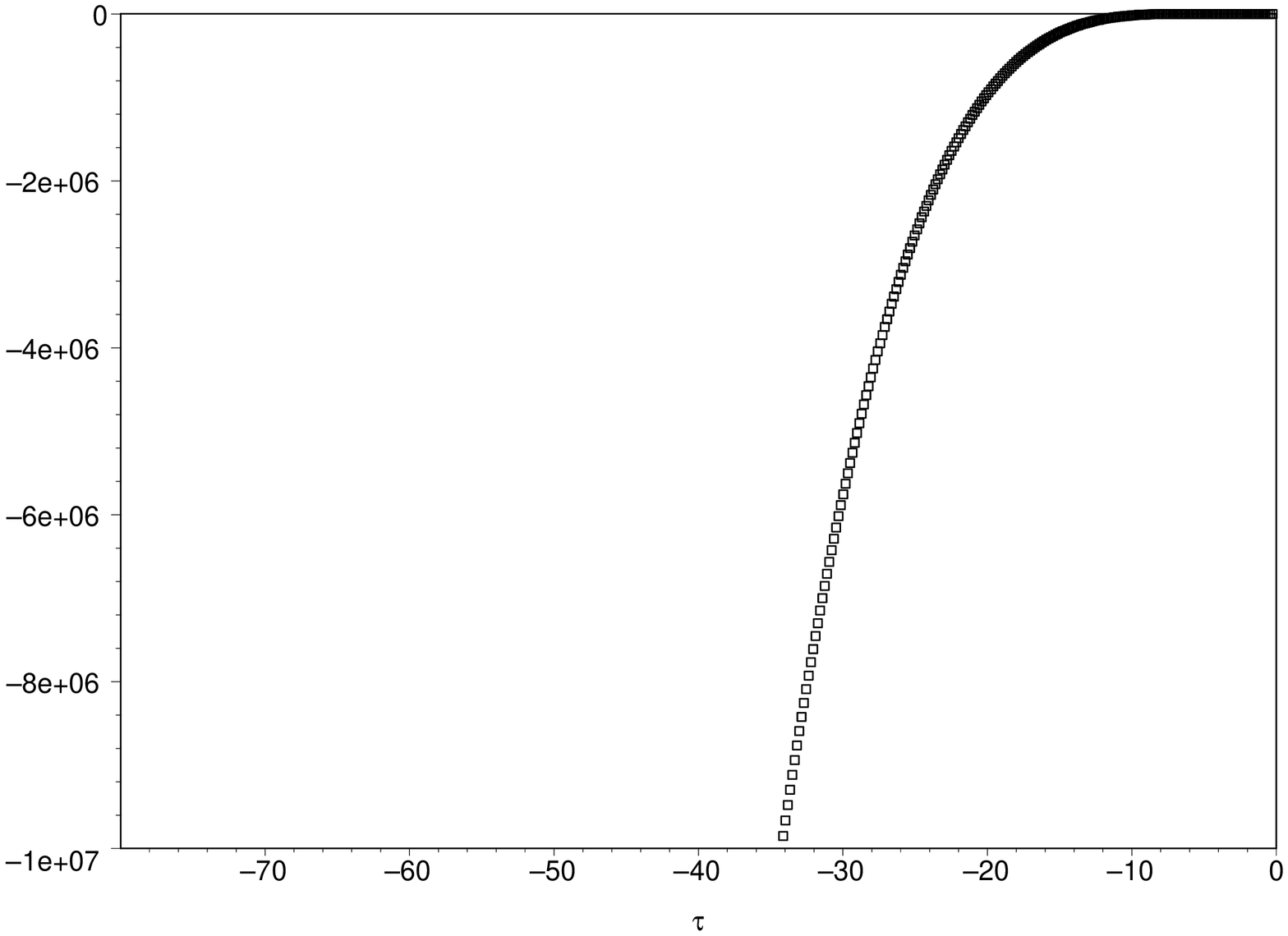}  &  \includegraphics[width=6cm,angle=270]{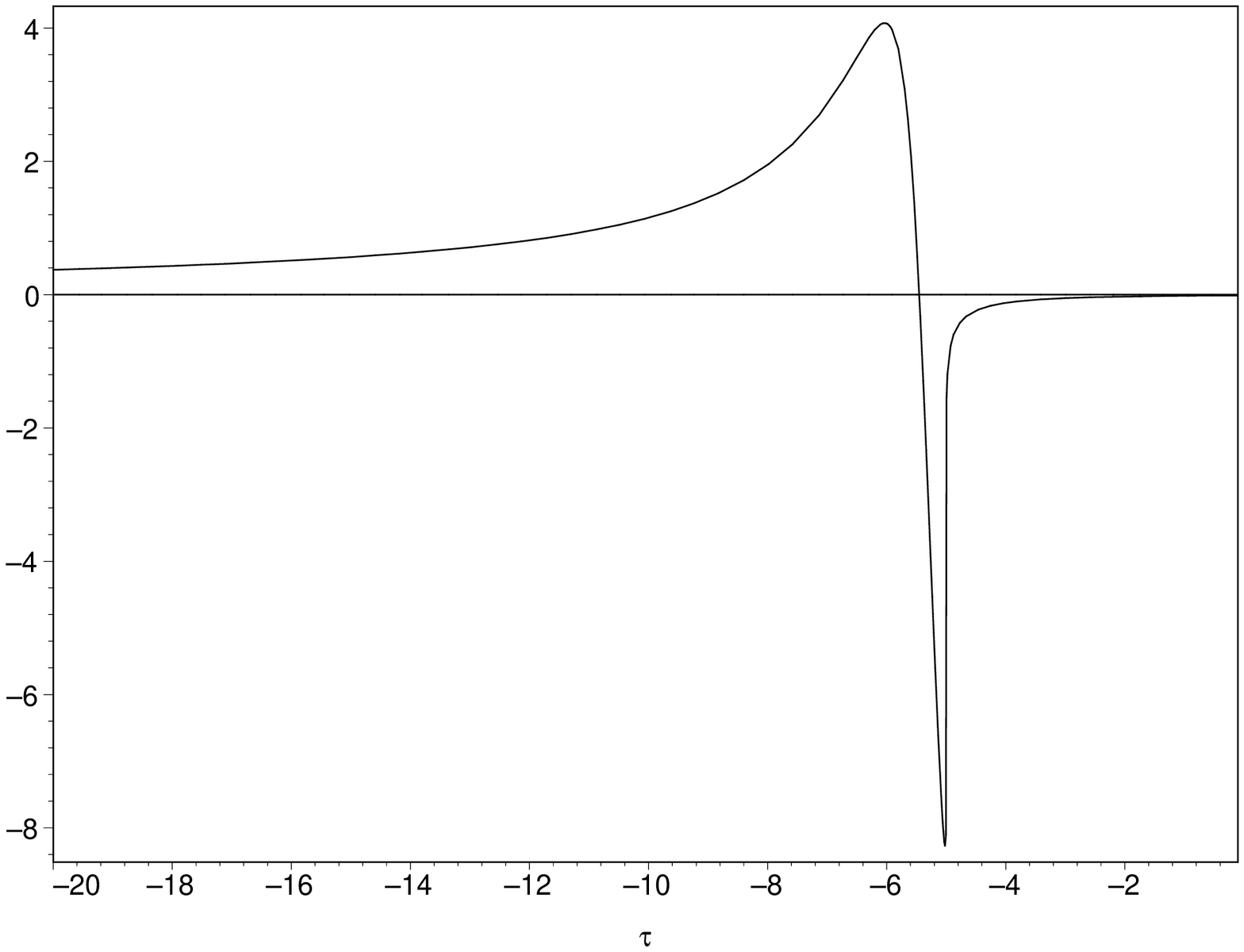}
\end{array}
$\caption{ 
 Top left: \emph{pump field} $M^2$  with neglected quantum corrections to the equation for tensor modes
 calculated numerically (circles) and with use of solutions (\ref{evol2}) (line).
 Top right : \emph{pump field} $M^2$ with quantum corrections to the equation for tensor modes (boxes)
 and without corrections (circles).
 Bottom left : Global behavior of the  \emph{pump field} $M^2$  with quantum corrections to equation for tensor modes.
 Bottom right : Evolution of the friction term $D'/D$ in the equation for tensor modes.
 In all panels it is assumed $\tau_0=5,\ j=100  \  \mbox{and} \ l=0.1$.}
\label{pumpfield}
\end{figure}

Introducing the new variable $\mu=ah$ and changing the time for conformal time 
we can rewrite equation (\ref{tensormodes1}) to the form
\begin{equation}
\mu''-\frac{D'}{D}\mu' +\left[-D\nabla^2 -\frac{a''}{a}+\frac{a'}{a}\frac{D'}{D} \right]\mu=0
\label{tensormodes2}
\end{equation}
Since the fluctuations considered have the quantum origin we must change the 
classical $\mu$ for the corresponding operator $\hat{\mu}$. The field $\hat{\mu}$ and conjugate momenta 
$\hat{\pi}$ can be decomposed for the Fourier modes according to 
\begin{eqnarray}
\hat{\mu}(\vec{x},\tau) =\frac{1}{2(2\pi)^{3/2}} \int d^3k \left\{ \hat{\mu}_{\vec{k}} e^{-i\vec{k}\cdot \vec{x}} +  
\hat{\mu}_{\vec{k}}^{\dagger} e^{i\vec{k}\cdot \vec{x}}  \right\}   \label{decomp1} \ ,      \\
\hat{\pi}(\vec{x},\tau) =\frac{1}{2(2\pi)^{3/2}} \int d^3k \left\{ \hat{\pi}_{\vec{k}} e^{-i\vec{k}\cdot \vec{x}} +  
\hat{\pi}_{\vec{k}}^{\dagger} e^{i\vec{k}\cdot \vec{x}}  \right\}.  \label{decomp2}        
\end{eqnarray}
where the relation of commutation $[\hat{\mu}(\vec{x},\tau),\hat{\pi}(\vec{x},\tau)] = i \delta^{(3)} (\vec{x}-\vec{y})$ 
is fulfilled. The equation for the Fourier modes is now
\begin{equation}
 \hat{\mu}_{\vec{k}}''-\frac{D'}{D}\hat{\mu}_{\vec{k}}' +D \left[ k^2 
-M^2 \right]\hat{\mu}_{\vec{k}} =0
\label{tensormodes3}
\end{equation}
where 
\begin{equation}
 M^2 =\frac{1}{D}\left(\frac{a''}{a}-\frac{a'}{a}\frac{D'}{D} \right) 
\label{pump field}
\end{equation}
is called the \emph{pump field}. In the classical limit ($D=1$) equation (\ref{tensormodes3}) assumes 
the known form
\begin{equation}
\hat{\mu}_{\vec{k}}'' +\left[k^2-\frac{a''}{a}\right]\hat{\mu}_{\vec{k}}=0.
\label{tensormodes4}
\end{equation}
Because it is impossible to solve equation (\ref{tensormodes3}) analytically we must investigate 
the effect of quantum corrections numerically. Because $D$ is always positive we can have amplifications 
of the tensor modes when $k^2<M^2$. The \emph{pump field} function was shown in Fig.~\ref{pumpfield}.
In the first panel (left  up) we draw \emph{pump field} $M^2$  with neglected quantum corrections 
calculated numerically and with the use of solution (\ref{evol2}). What we see is that the 
numerically calculated \emph{pump field} extends the region of super-adiabatic amplifications.
In the next panel (left right) we compare the numerically calculated \emph{pump field}
with and without quantum corrections. We see that quantum corrections lower the 
region of amplification. The obtained value is however still larger than this obtained 
using the approximated analytical solution (\ref{evol2}).  

The friction term $D'/D$ in equation (\ref{tensormodes3}) can be positive or negative leading to the 
amplification or to the damping. We see that the friction term $D'/D$ is for the most of time positive
leading to amplification and drastically peak to negative values in the neighborhood of $a_*$. In fact this 
dependence strongly depends of a quantum number $l$. The increasing value of $l$ the negative pick goes  
toward to positive values and becomes less sharp. 

We can now use the approximation $D=1$ in equation (\ref{tensormodes3}) to calculate the spectrum of tensor perturbations
during the super-inflationary phase. The spectrum of tensor perturbations can be now expressed using correlation function
\begin{equation}
\langle 0|  \hat{h}^i_j(\vec{x},\tau) \hat{h}^j_i (\vec{y},\tau)| 0 \rangle =
\frac{64\pi G}{a^2} \int \frac{d^3k}{(2\pi)^3} |\hat{\mu}_{\vec{k}}(\tau)|^2 e^{-i\vec{k}\cdot\vec{r}} \equiv 
\int \frac{dk}{k} \mathcal{P}_T(k) \frac{\sin kr}{kr}   \label{spectdeff}
\end{equation}
where an Einstein convention of summation was used on the left side.  
For considered super-inflationary phase, using expression (\ref{evol1}), we have solution 
\begin{equation}
\mu_{k} =\frac{\mathcal{N}}{\sqrt{2k}} \sqrt{-k\tau} H^{(1)}_{\beta+\frac{1}{2}} (-k\tau)
\end{equation}
where 
\begin{equation}
\mathcal{N} = \sqrt{\frac{\pi}{2}} e^{i\pi (\nu+1/2)/2} \ \ \mbox{and} \ \ \beta=2\frac{1-l}{2+l}
\end{equation}
Normalization is found by correspondence to well normalized plane wave $e^{-ik\tau}/\sqrt{2k}$ 
for high energetical modes $|k\tau|\ll 1$.
Since for us interesting are super-horizontal modes we can use approximation
\begin{equation}
H^{(1)}_{\nu}(-k\tau) \simeq  -\frac{i}{\pi} \Gamma (\nu) \left(  -\frac{k\tau}{2} \right)^{-\nu}.
\end{equation}
Super-horizontal modes are these which firstly cross out the horizon and then evolve ''frozen'' in super-horizontal scales.
Finally, in further epochs on universe, such a modes reenter horizon. These modes bring an information from earliest 
stages of the universe. Given modes cross the horizon when     
\begin{equation}
k \simeq aH = \frac{\beta}{\tau_0}\left( \frac{a}{a_{*}} \right)^{\frac{1}{\beta}}  ,
\end{equation}
where we used  definition of Hubble radius  and equation of evolution (\ref{evol1}) . 
Finally with use of definition  (\ref{spectdeff})  the spectrum at horizon crossing have a form
\begin{equation}
\mathcal{P}_T(k) = \mathcal{A}_T^2 k^{n_{T}}
\end{equation}
where spectral index is equal
\begin{equation}
n_{T} = \frac{d \ln \mathcal{P}_T }{d \ln k}   = \frac{6l}{2+l}
\label{spectind}
\end{equation}
and normalization constant is expressed as
\begin{equation}
\mathcal{A}_T^2 = \frac{ \Gamma^2(\beta+1/2) 2^{2\beta+4}}{m_{\text{Pl}}^2\pi^{2} a^2_{*} } 
\left(\frac{\beta}{\tau_0}\right)^{2\beta}  .
\end{equation}
In this case tensor spectral index (\ref{spectind}) is positive and $n_{T}\in (0,3)$. To compare, from the standard slow-roll
inflation tensor spectral index is $n_T=-2\epsilon+\mathcal{O}(\epsilon^2)$. Unfortunately available data from CMB or
from large scale structures observations are not sufficiently precise to determinate value of the tensor spectral index.

\section{Relic gravitons from the quantum to classical universe transition}

In the previous section we derived and investigated equation for tensor modes (\ref{tensormodes3}).  
Using approximations we also calculated the spectrum of gravitons which cross the 
horizon during the super-inflationary phase. Now we want to calculate a number 
of gravitons which are produced during the transition from the quantum to 
classical regime of evolution. Before we start it, let us calculate the width 
of the band of produced gravitons. It can be directly taken from condition $k^2<M^2$.
We use the classical approximation for the evolution of tensor modes for it preserves 
the physical picture of the graviton creation process as the previous numerical 
investigations indicate. In this approximation ($D=1$) we have a maximum of frequency 
for $\tau_0$, so with the use of equation (\ref{evol1}) we have  
\begin{equation}
k_{\text{max}}=  \sqrt{\beta(\beta+1)} \frac{1}{\tau_{0}}. 
\label{kmax}
\end{equation}
In fact, as it can be seen in Fig.~\ref{pumpfield}, this frequency is generally higher. 
The corresponding maximal frequency for the present epoch is
\begin{equation}
\nu_{\text{max}}=\frac{k_{\text{max}}}{2\pi a_*} \left( \frac{a_*}{a_{\text{today}}} \right)=
\frac{\sqrt{\beta(\beta+1)}}{2\pi a_*} \frac{1}{\tau_{0}}  \left( \frac{a_*}{a_{\text{today}}} \right).
\label{numax}
\end{equation}    
To estimate this value we can approximate
\begin{equation}
\frac{a_{\text{today}}}{a_*} \sim \frac{T_{\text{Pl}}}{T_{\text{CMB}}} = 
\frac{1.4 \cdot 10^{32} \ \text{K}}{3.7 \ \text{K}} \simeq 10^{32} 
\end{equation}
where $T_{\text{Pl}}$ is the Planck temperature. The other way to estimate 
value ${a_{\text{today}}}/{a_*}$ is to use the Friedmann equation with
radiation. This gives equation
\begin{equation}
\frac{a_{\text{today}}}{a_*} = \sqrt{\frac{H_{today}}{H_*}} = 
\sqrt{\frac{H_{today}t_{\text{Pl}}\tau_0}{\beta} \sqrt{\frac{\gamma j }{3}} }   
\label{aa}
\end{equation}
where we use solution (\ref{evol2}) to calculate $H_*$. To obtain a numerical 
value we must know $\tau_0$. We use here the constraint for energy in the form 
$|\dot{\phi}_i|/m_{\text{Pl}}^2 < 1$ \cite{Mulryne:2005ef} (the kinetic energy 
dominates over the contribution from the potential part as we mentioned before) 
for $a_i=\sqrt{\gamma}l_{\text{Pl}}$. Below the value of chosen $a_i$ the space 
becomes discrete and the smooth dynamical equations cannot be used. The boundary
for the kinetic energy is introduced to prevent energies beyond the Planck scale 
being produced. With use of this boundary conditions and the Friedmann equation 
(\ref{Fried1}) with solution (\ref{evol2}) we obtain the constraint for the 
conformal time $\tau_0$   
\begin{equation}
\tau_0 >  \frac{1-l}{2+l} \sqrt{\frac{3}{\pi \gamma}} \left( \frac{3}{1+l} \right)^{\frac{3}{2}\frac{1}{2-2l}}
\left(\frac{3}{j} \right)^{\frac{1}{2}\frac{4-l}{1-l}}. 
\label{bound1}
\end{equation}
As an exemple for the model with $l=0.1$ and $j=100$ we obtain $\tau_0 > 0.0014 $ and 
for the model with $l=3/4$ and $j=100$ we obtain $\tau_0 > 1.6 \cdot 10^{-8} $. We see that this boundary
depends very strongly on the quantum numbers. Combining equation (\ref{numax}) with (\ref{aa}) we see that
$\nu_{\text{max}}\propto \tau_{0}^{-1/2}$, so the boundary (\ref{bound1}) gives us also the upper constraint for
a maximal value of frequency  $\nu_{\text{max}}$.  For the model with $l=0.1$ and $j=100$ we have
$\nu_{\text{max}} < 6.6 \cdot 10^{14} $ Hz and for the model with $l=3/4$ and $j=100$ we obtain 
$\nu_{\text{max}} < 2.8 \cdot 10^{24} $ Hz.
Generally values of $\nu_{\text{max}}$ can be smaller than boundary values.
For the further studies we choose the model with $\tau_0=0.1$.  
So in this case the width of the band of relic gravitons considered is nowadays
$[0, 10^3 \ \mbox{GHz} ] $ for $l=0.1$. 

Fourier modes of (\ref{decomp1}) and 
(\ref{decomp2}) for the super-inflationary evolution (\ref{evol1}) can be 
written with the use of annihilation and creation operators as follow
\begin{eqnarray}
\hat{\mu}_{\vec{k}}(\tau)  &=&  \hat{a}_{\vec{k}} f_1(k,\tau)+ 
\hat{a}_{-\vec{k}}^{\dagger} f_1^{*}(k,\tau)  
 \ \ \mbox{for}   \ \  \tau<-\tau_0 \ ,  \\
\hat{\pi}_{\vec{k}}(\tau)  &=&  \hat{a}_{\vec{k}} g_1(k,\tau)+ 
\hat{a}_{-\vec{k}}^{\dagger} g_1^{*}(k,\tau)  
  \ \ \mbox{for}   \ \  \tau<-\tau_0 .
\end{eqnarray}
In this case the values of coefficients are 
\begin{eqnarray}
f_1(k,\tau)   &=& \frac{\mathcal{N}_1}{\sqrt{2k}} \sqrt{-k\tau} H^{(1)}_{\nu}(-k\tau) \ ,    \\
g_1(k,\tau)   &=&- \mathcal{N}_1\sqrt{\frac{k}{2}} \sqrt{-k\tau} \left[ - H^{(1)}_{\nu+1}(-k\tau) +\frac{1+2\nu}{2(-k\tau)} 
H^{(1)}_{\nu}(-k\tau)       \right] 
\end{eqnarray}
where 
\begin{equation}
\mathcal{N}_1 = \sqrt{\frac{\pi}{2}} e^{i\pi (\nu+1/2)/2} \ \ \ \mbox{and} \ \ \ \nu=\beta+\frac{1}{2}  .
\end{equation}
Similarly, modes of (\ref{decomp1}) and (\ref{decomp2}) for the classical 
evolution (\ref{evol2}) we can be written down as
\begin{eqnarray}
\hat{\mu}_{\vec{k}}(\tau)  &=&  \hat{b}_{\vec{k}} f_2(k,\tau)+ \hat{b}_{-\vec{k}}^{\dagger} f_2^{*}(k,\tau)
  \ \ \mbox{for}    \ \  \tau>-\tau_0   \ ,  \\
\hat{\pi}_{\vec{k}}(\tau)  &=&  \hat{b}_{\vec{k}} g_2(k,\tau)+ \hat{b}_{-\vec{k}}^{\dagger} g_2^{*}(k,\tau) 
\ \ \mbox{for}    \ \  \tau>-\tau_0  \ .
\end{eqnarray}
Where the coefficients of decomposition are 
\begin{eqnarray}
f_2(k,\tau)   &=& \mathcal{N}_2  \sqrt{1+4\frac{1-l}{2+l}\left(\frac{\tau_0+\tau}{\tau_0}\right) }  H_0^{(2)} 
\left(k\tau +k\zeta    \right)    \exp{\left(       ik\zeta  \right) }  \ ,     \\
g_2(k,\tau)   &=& \frac{\mathcal{N}_2}{ \tau_0}\left[\frac{H_0^{(2)} 
\left(k\tau +k\zeta    \right) }{ \sqrt{1+4\frac{1-l}{2+l}\left(\frac{\tau_0+\tau}{\tau_0}\right) }}\frac{2(1-l)}{2+l}-k\tau_0  
\sqrt{1+4\frac{1-l}{2+l}\left(\frac{\tau_0+\tau}{\tau_0}\right) } H_1^{(2)} 
\left(k\tau +k\zeta   \right)  \right] \exp{\left( ik\zeta  \right) }
\end{eqnarray}
with
\begin{equation}
\mathcal{N}_2 = \frac{\sqrt{\pi}}{4}\sqrt{\tau_0} \sqrt{ \frac{2+l}{1-l} } e^{-i\pi/4} \ \ \ \mbox{and} \ \ \
\zeta = \tau_0\frac{3}{4}\frac{2-l}{1-l}
\end{equation}
where $H^{(2)}$ is the Haenkel function of the second kind.
 
\begin{figure}[ht!]
\centering
$\begin{array}{cc}   
\includegraphics[width=6cm,angle=270]{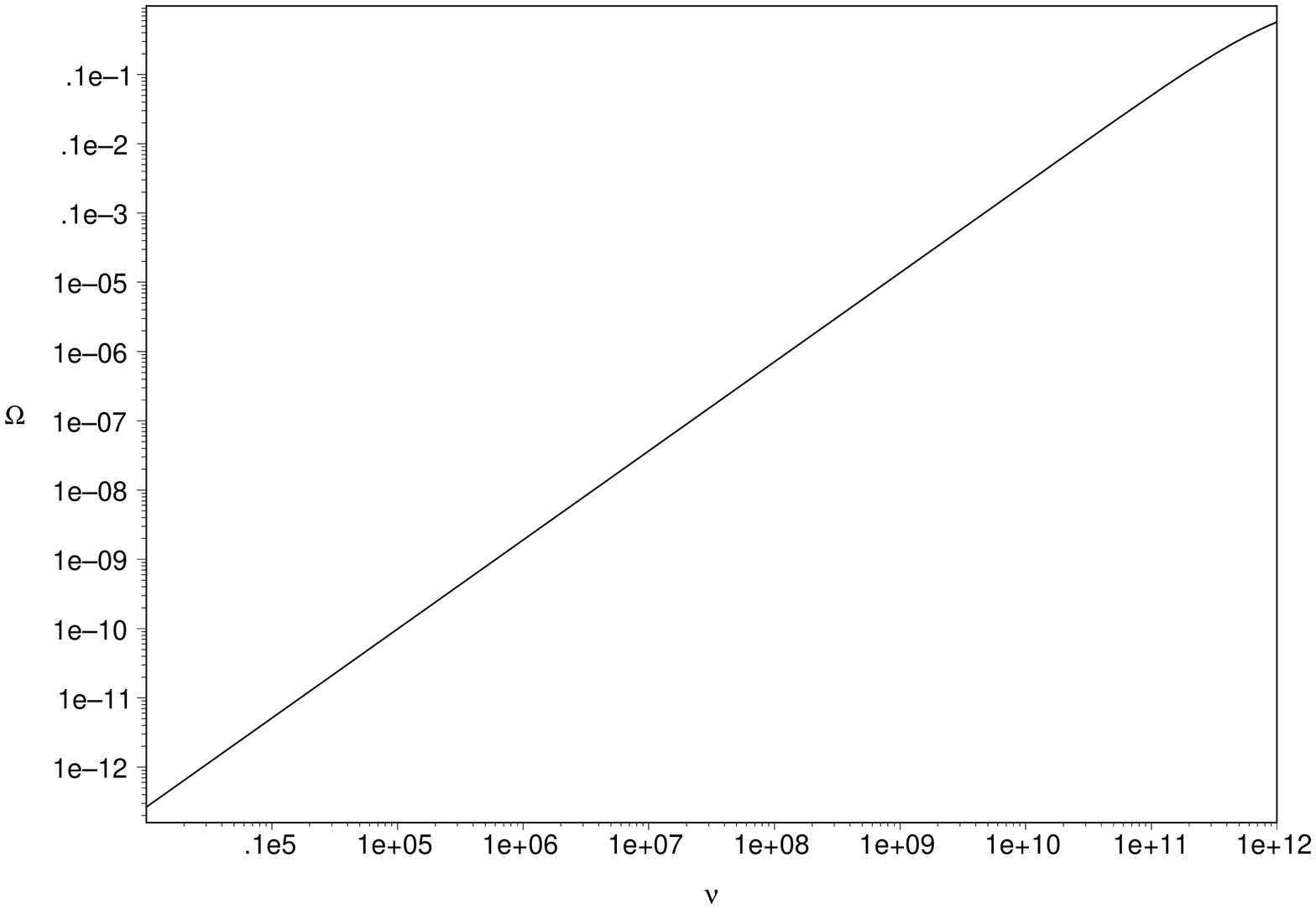}  &  \includegraphics[width=6cm,angle=270]{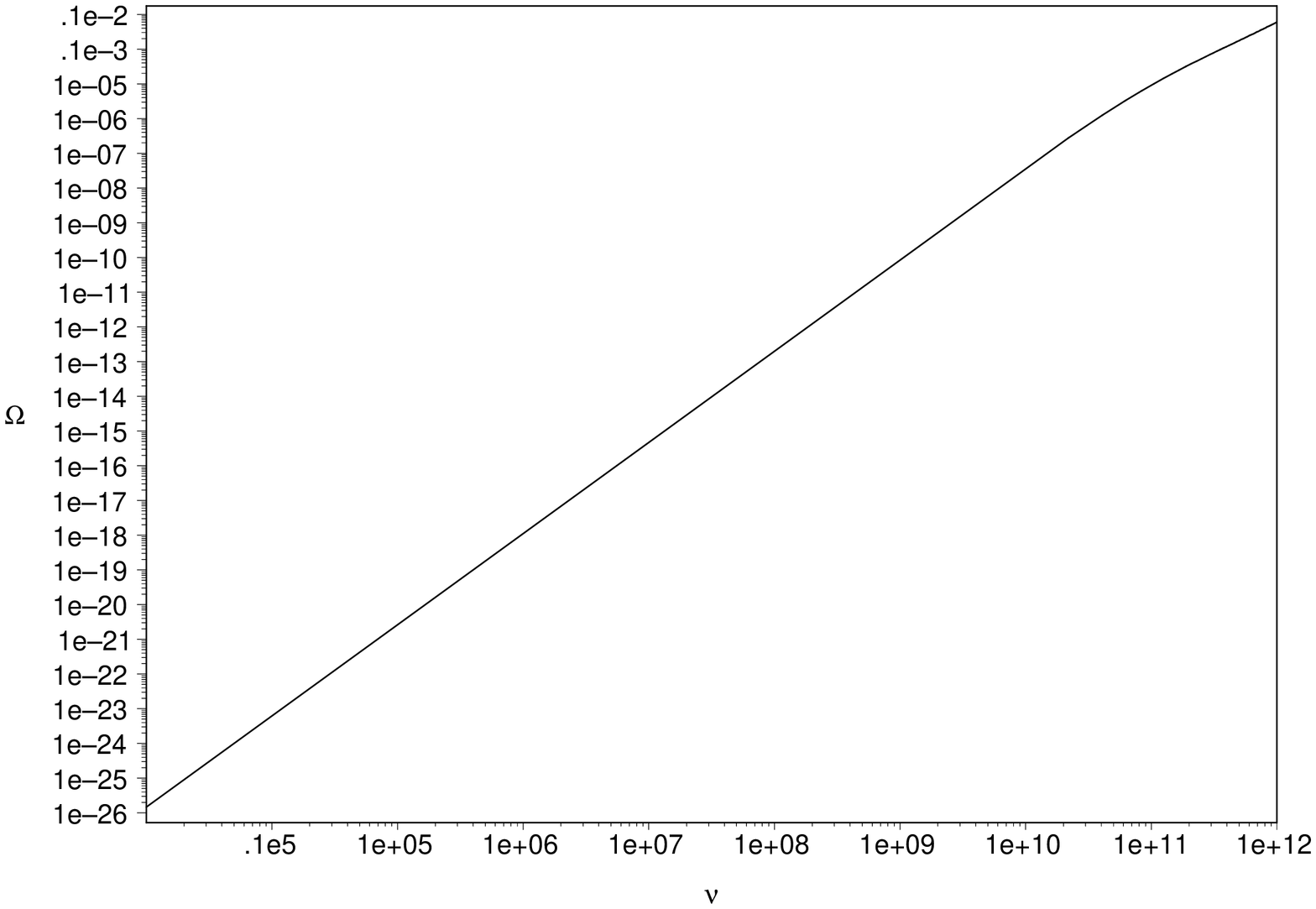}   
\end{array}
$\caption{ Left: Function $\Omega_{\text{gw}}(\nu)$ with $ j=100, \tau_0=0.1  \  \mbox{and} \ l=0.1$.
           Right: Function $\Omega_{\text{gw}}(\nu)$ with $ j=100, \tau_0=0.1  \  \mbox{and} \ l=3/4$. Frequency scales in Hertz.}
\label{pict01}
\end{figure}

The main idea of particles creation during transition comes from the Bogoliubov transformation
\begin{eqnarray}
\hat{b}_{\vec{k}} &=& B_{+}(k) \hat{a}_{\vec{k}} + B_{-}(k)^{*}  \hat{a}_{-\vec{k}}^{\dagger} \ , \label{Bog1}  \\
\hat{b}_{\vec{k}}^{\dagger} &=& B_{+}(k)^{*}\hat{a}_{\vec{k}}^{\dagger} + B_{-}(k)    \hat{a}_{-\vec{k}} \label{Bog2} 
\end{eqnarray}
where from relations of commutation $[\hat{a}_{\vec{k}},\hat{a}_{\vec{k}}^{\dagger}]=\delta^{(3)}(\vec{k}-\vec{p})$ 
and $[\hat{b}_{\vec{k}},\hat{b}_{\vec{k}}^{\dagger}]=\delta^{(3)}(\vec{k}-\vec{p})$  we have $|B_{+}|^2-|B_{-}|^2=1$.
In the quantum phase we have $\hat{a}_{\vec{k}}|0_{in}\rangle=0$ where $|0_{in}\rangle$ is the vacuum state of this phase. 
In the final classical epoch, similarly $\hat{b}_{\vec{k}}|0_{out}\rangle=0$ what defines the new vacuum state
$|0_{out}\rangle$. But since we are in the Heisenberg description the true vacuum state in the
classical phase is $|0_{in}\rangle$ and thanks to the mixing from the Bogoliubov transformation (\ref{Bog1}) 
we have $ \hat{b}_{\vec{k}}|0_{in}\rangle =
B_{-}(k)^{*}  \hat{a}_{-\vec{k}}^{\dagger}|0_{in}\rangle$. So when $B_{-}(k)$ is the nonzero coefficient we have
the production of particles (gravitons) in the final state. What we need now is to calculate coefficients of the 
Bogoliubov transformation $B_{-}(k)$ and $B_{+}(k)$  which can be written as 
\begin{eqnarray}
B_{-}(k)&=&\frac{f_1(-\tau_0)g_2(-\tau_0) -g_1(-\tau_0) f_2(-\tau_0)}{f_2^*(-\tau_0)g_2(-\tau_0)-g_2^*(-\tau_0)f_2(-\tau_0)} \ , 
 \\
B_{+}(k)&=& \frac{f_1(-\tau_0)g_2^*(-\tau_0) -g_1(-\tau_0) f_2^*(-\tau_0)}{f_2(-\tau_0)g_2^*(-\tau_0)-g_2(-\tau_0)f_2^*(-\tau_0)}.
\end{eqnarray}
Since the total momentum of produced gravitons is conserved we can write the 
expression for the number of produced particles
\begin{equation}
\bar{n}_{\vec{k}} = \frac{1}{2} \langle 0_{in} |[ \hat{b}_{\vec{k}}^{\dagger}\hat{b}_{\vec{k}}+
 \hat{b}_{-\vec{k}}^{\dagger}\hat{b}_{-\vec{k}} ]| 0_{in} \rangle =|B_{-}(k)|^2.
\end{equation}
As we can see, to calculate a number of gravitons we only need to know the 
coefficient $B_{-}(k)$. Now we can calculate the function 
$\Omega_{\text{gw}}(\nu)$ defined in equation (\ref{omegaGW}). The essential 
energy density is from the relation 
\begin{equation}
d\rho_{\text{gw}} = 2 \cdot \hslash \omega \cdot  \frac{4\pi \omega^2 d\omega}{(2\pi c)^3} \cdot \bar{n}_{\vec{k}}
\end{equation}
where factor 2 comes from two polarizations of gravitational waves. With the 
use of relation (\ref{numax}) we finally obtain the equation
\begin{equation}
\Omega_{\text{gw}}(\nu) = 3.7 \cdot 10^{-49} h^{-2}_{0} \nu^4 \ \bar{n}\left( \sqrt{\beta(\beta+1)}
 \frac{\nu}{\nu_{\text{max}}} \right) 
\end{equation}
where $h_0$ is the normalized Hubble rate $h_0=H_0/100 \ \mbox{km}^{-1}\ \mbox{s} \ \mbox{Mpc}$.
We compute this function and show it in the logarithmic plot with $l=0.1$ and $l=3/4$ 
(Fig.~\ref{pict01}). This spectrum is characterized by 
a hard branch with the maximum for $\sim 10^{12} \ \mbox{Hz}$ for $l=0.1$ and $\sim 10^{11} \ \mbox{Hz}$  for $l=3/4$.
In this limit $\Omega_{\text{gw}}$ approaches respectively to $\sim 10^{-1}$  and $\sim 10^{-5}$. 

\begin{figure}[ht!]
\centering
\includegraphics[width=7cm,angle=270]{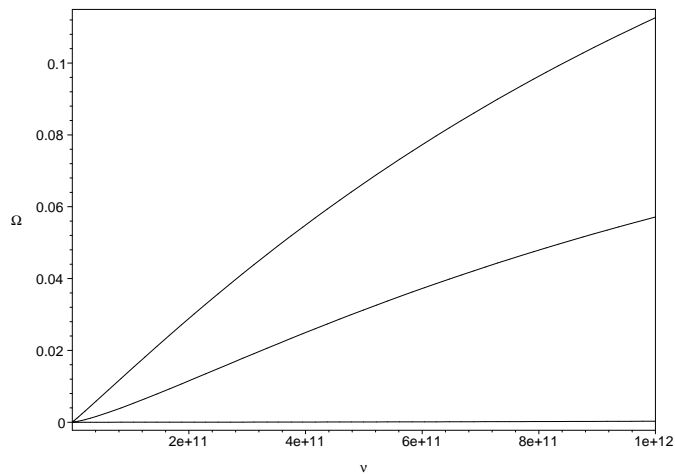}
\caption{ Function $\Omega_{\text{gw}}(\nu)$   for $l=0.01,0.1,3/4 $ (from top to bottom), $\tau_0=0.1 \  \text{and} \ j=100$.
  Frequency scale in Hertz.}
\label{pict001-01-34}
\end{figure}

When the high energy region is shown only, the dependence 
$\Omega_{\text{gw}}(\nu)$ on the quantum parameter $l=0.01,0.1,3/4 $ is 
exhibited (Fig.~\ref{pict001-01-34}).

As we mentioned in section {\bf I}, recent constraints from LIGO are 
$\Omega_{\text{gw}} < 6.5 \cdot 10^{-5}$ \cite{Abbott:2006zx}. The LIGO 
observations are however concentrated in the region of $\sim 10^2 \ \mbox{Hz}$. 
From Loop Quantum Cosmology we have in this region $\Omega_{\text{gw}}\sim 10^{-14}$ 
(for $l=0.1$), what is extremely below the observational sensitivity. The 
numerical values obtained by us contain estimations of the time of transition 
to classical universe. The used value should be somehow proper to the order of 
magnitude. So we expect also similar deviations of $\Omega_{\text{gw}}(\nu)$.

The spectrum obtained here is not a distinct feature of Loop Quantum Cosmology.
As it was shown by Giovannini \cite{Giovannini:1999bh} a similar high energy 
branch was obtained in the quintessential inflationary model. Calculations 
based on String Cosmology lead also to similar results \cite{Brustein:1995ah}. 
To compare, for the standard inflationary models the spectrum is flat.

\section{Summary}

Loop Quantum Cosmology has received much attention in the theoretical 
astrophysics. But what was lacked so far was empirical consideration of this 
theory. Bojowald indicated the quantum effects are negligible small at the 
present epoch but they can potentially tested \cite{Bojowald:2007ab}.
Along Bojowald's lines we showed that gravitational waves can be the 
real observable for testing Loop Quantum Gravity effects. 

We have considered the transition from the semi-classical to classical universe 
described by Loop Quantum Cosmology. 
In the analytical approximation we obtained the tensor energy spectrum of 
the relic gravitons from the super-inflationary phase.
The analytical model takes into consideration the corrections to 
dynamical evolution only. While taking corrections to the equation for 
the tensor modes this equation cannot be solved analytically, so it is
only possible to consider it numerically.
The numerical investigation of the equation for tensor modes
gave us that lower $\nu_{\text{max}}$ is admissible when the 
loop quantum effects are incorporated. The loop quantum
gravity effects product additional damping during the production 
of gravitons. This is a challenge for future investigation - the full numerical 
analysis of this model.
  
When we considered the production of gravitons during the transition phase the spectrum
of these gravitons is characterized by the hard branch. The corresponding value of the 
parameter $\Omega_{\text{gw}}$, in its maximum, is 
$\Omega_{\text{gw}} \sim 10^{-7} \dots 10^{-1}$, depending on the value of the 
parameter of quantization $l$. In the region of the LIGO highest sensitivity 
we obtained the very small value of the parameter $\Omega_{\text{gw}}$, namely 
$\sim 10^{-14}$ for $l=0.1$ and $\sim 10^{-28}$ for $l=3/4$. As we mentioned, 
the similar hard branch is also a feature of 
quintessential inflationary and String Cosmology models. This work gives the 
motivation to search for high energetic gravitational waves.

\begin{acknowledgments}
This work was supported in part by the Marie Curie Actions Transfer of
Knowledge project COCOS (contract MTKD-CT-2004-517186). The authors are 
grateful to the members of the seminar on observational cosmology for 
discussion and comments, especially dr Adam Krawiec. We would like also to 
thank the anonymous referee for important remarks. 
\end{acknowledgments}


\begin{thebibliography}{99}

\bibitem{Bojowald:2001xe}
  M.~Bojowald,
  Phys.\ Rev.\ Lett.\  {\bf 86} (2001) 5227
  [arXiv:gr-qc/0102069].

\bibitem{Bojowald:2005zk}
  M.~Bojowald,
  J.\ Phys.\ Conf.\ Ser.\  {\bf 24} (2005) 77
  [arXiv:gr-qc/0503020].

\bibitem{Stachowiak:2006uh}
  T.~Stachowiak and M.~Szydlowski,
  Phys.\ Lett.\  B {\bf 646} (2007) 209
  [arXiv:gr-qc/0610121].

\bibitem{Bojowald:2002nz}
  M.~Bojowald,
  Phys.\ Rev.\ Lett.\  {\bf 89} (2002) 261301
  [arXiv:gr-qc/0206054].

\bibitem{Tsujikawa:2003vr}
  S.~Tsujikawa, P.~Singh and R.~Maartens,
  Class.\ Quant.\ Grav.\  {\bf 21} (2004) 5767
  [arXiv:astro-ph/0311015].

\bibitem{Hossain:2005}
G.~M.~Hossain,
Class.\ Quant.\ Grav.\ {\bf 22}, 2511 (2005)
[arXiv:gr-qc/0411012].

\bibitem{Calcagni:2007}
G.~Calcagni and M.~Cortes,
Class.\ Quant.\ Grav.\ {\bf 24}, 829 (2007)
[arXiv:gr-qc/0607059].

\bibitem{Mulryne:2006}
D.~J.~Mulryne and N.~J.~Nunes,
Phys.\ Rev.\ D {\bf 74}, 083507 (2006)
[arXiv:astro-ph/0607037].

\bibitem{Ashtekar:1997yu}
  A.~Ashtekar, J.~Baez, A.~Corichi and K.~Krasnov,
  Phys.\ Rev.\ Lett.\  {\bf 80} (1998) 904
  [arXiv:gr-qc/9710007].

\bibitem{Abbott:2003vs}
  B.~Abbott {\it et al.}  [LIGO Scientific Collaboration],
  Nucl.\ Instrum.\ Meth.\  A {\bf 517} (2004) 154
  [arXiv:gr-qc/0308043].

\bibitem{Abbott:2007wd}
  B.~Abbott {\it et al.}  [ALLEGRO Collaboration],
  arXiv:gr-qc/0703068.

\bibitem{Abbott:2006zx}
  B.~Abbott {\it et al.}  [LIGO Scientific Collaboration],
  Astrophys.\ J.\  {\bf 659} (2007) 918
  [arXiv:astro-ph/0608606].

\bibitem{Bojowald:2006da}
  M.~Bojowald,
  Living Rev.\ Rel.\  {\bf 8} (2005) 11
  [arXiv:gr-qc/0601085].

\bibitem{Bojowald:2004ax}
  M.~Bojowald,
  Pramana {\bf 63} (2004) 765
  [arXiv:gr-qc/0402053].

\bibitem{Bojowald:2002ny}
  M.~Bojowald,
  Class.\ Quant.\ Grav.\  {\bf 19} (2002) 5113
  [arXiv:gr-qc/0206053].

\bibitem{Giovannini:2004rj}
  M.~Giovannini,
  Int.\ J.\ Mod.\ Phys.\  D {\bf 14} (2005) 363
  [arXiv:astro-ph/0412601].

\bibitem{Giovannini:2007xh}
  M.~Giovannini,
  arXiv:astro-ph/0703730.

\bibitem{Mulryne:2005ef}
  D.~J.~Mulryne, R.~Tavakol, J.~E.~Lidsey and G.~F.~R.~Ellis,
  Phys.\ Rev.\  D {\bf 71} (2005) 123512
  [arXiv:astro-ph/0502589].

\bibitem{Giovannini:1999bh}
  M.~Giovannini,
  Phys.\ Rev.\  D {\bf 60} (1999) 123511
  [arXiv:astro-ph/9903004].

\bibitem{Brustein:1995ah}
  R.~Brustein, M.~Gasperini, M.~Giovannini and G.~Veneziano,
  Phys.\ Lett.\  B {\bf 361} (1995) 45
  [arXiv:hep-th/9507017].

\bibitem{Bojowald:2007ab}
  M.~Bojowald,
  arXiv:gr-qc/0701142.

\end{thebibliography}
\end{document}